\documentclass[a4paper]{jpconf}
\usepackage{graphicx}
\usepackage{graphicx,pgfarrows,pgfnodes,amssymb}
\usepackage{amsmath, amscd, amsthm, amssymb, mathrsfs,amsfonts}
\usepackage{tikz,ulem}
\usepackage[all,knot]{xy}
\usepackage{chngcntr}
\counterwithout{figure}{section}\usepackage{hyperref}

\def\hrad{0.5}
\def\vrad{0.1}
\def\rsep{0.25}
\def\miny{0}
\def\maxy{4}
\def\tension{0.5}

\newcommand{\C}{\mathbb C}

\newcommand{\Z}{\mathbb Z}
\newcommand{\Q}{\mathbb Q}

\newcommand{\B}{\mathcal{B}}
\newcommand{\cH}{\mathcal{H}}

\newcommand{\LL}{\mathcal{L}}

\newcommand{\one}{\mathbf{1}}

\newtheorem{thm}{Theorem}[section]

\newtheorem{conj}[thm]{Conjecture}

\newtheorem{exercise}[thm]{Exercise}
\newtheorem{example}[thm]{Example}

\begin{document}

\title{An Invitation to the Mathematics of Topological Quantum Computation}

\author{Eric C. Rowell}

\address{Mathematics Department, Texas A\&M University, College Station, TX 77843-3368}

\ead{rowell@math.tamu.edu}

\begin{abstract}
Two-dimensional topological states of matter offer a route to quantum computation
that would be topologically protected against the nemesis of the quantum circuit model:
decoherence. Research groups in industry, government and academic institutions are pursuing this approach. We give a mathematician's perspective
on some of the advantages and challenges of this model, highlighting some recent advances.  We then give a short description of how we might extend the theory
to three-dimensional materials.
\end{abstract}

\section{Introduction}

In \cite{FKLW} we find the following convenient definition:
\textit{\textbf{Quantum computation} is any computational model based upon
the theoretical ability to manufacture, manipulate and measure
quantum states.} A topological quantum computer is a hypothetical device that relies upon a kind of topological symmetry in topologically ordered states of matter to carry out fault-tolerant quantum computation.  Usually, these topological phases are taken to be (effectively) 2-dimensional systems of anyons: point-like quasi-particles that emerge in certain condensed matter systems.
Topological phases of matter were first realized through quantum Hall effects, for example in fractional quantum Hall liquids in the experiments of Tsui and St\"ormer in 1982 \cite{TsuiStormer} which led to the Nobel prize they shared with Laughlin \cite{Laugh} in 1998.
In the past few years \cite{Kit,Fr,FKLW} the possibility of building a quantum computer using physical systems exhibiting topological phases has motivated significant investment of resources towards realizing such a project.  Both Microsoft and Alcatel-Lucent are currently pursuing topological qubits.  Besides the potential commercial computational benefits, these states of matter are fascinating from both the physical and mathematical perspectives, connecting the seemingly unrelated subjects of quantum topology and condensed matter.

The mathematical foundations of topological quantum computation are usually cast in categorical language.  In this survey we will avoid categorical terminology in the hopes that this will broaden the readership, without significantly sacrificing precision.  There are other expository works of varying length and precision on the subject.  The shortest and most elementary is \cite{Collins}, written for a general audience.  The text \cite{Pachos} gives a fairly comprehensive account aimed at physicists, while the excellent survey \cite{Nayaketal} mainly focuses on topological and condensed matter themes.  The short survey \cite{FKLW} introduces the topological model to mathematicians,
while the very precise and complete \cite{WangBook} makes full use of the categorical language.

The main goal of this survey is to give non-experts a taste of the mathematics of topological quantum computation, illustrating how the theory arises naturally and hopefully inspiring further reading.
In the first part of this survey we will present a first approximation of the mathematical model for anyons on surfaces.  Some technical details will be neglected, but these can be reconciled by further reading (for example \cite{WangBook}).  The second part will be devoted to describing some of the recent advances and open problems based upon this model.

\section{Anyons on surfaces}

\subsection{Motivation}
\begin{figure}[h]
 \includegraphics[width=3.5in]{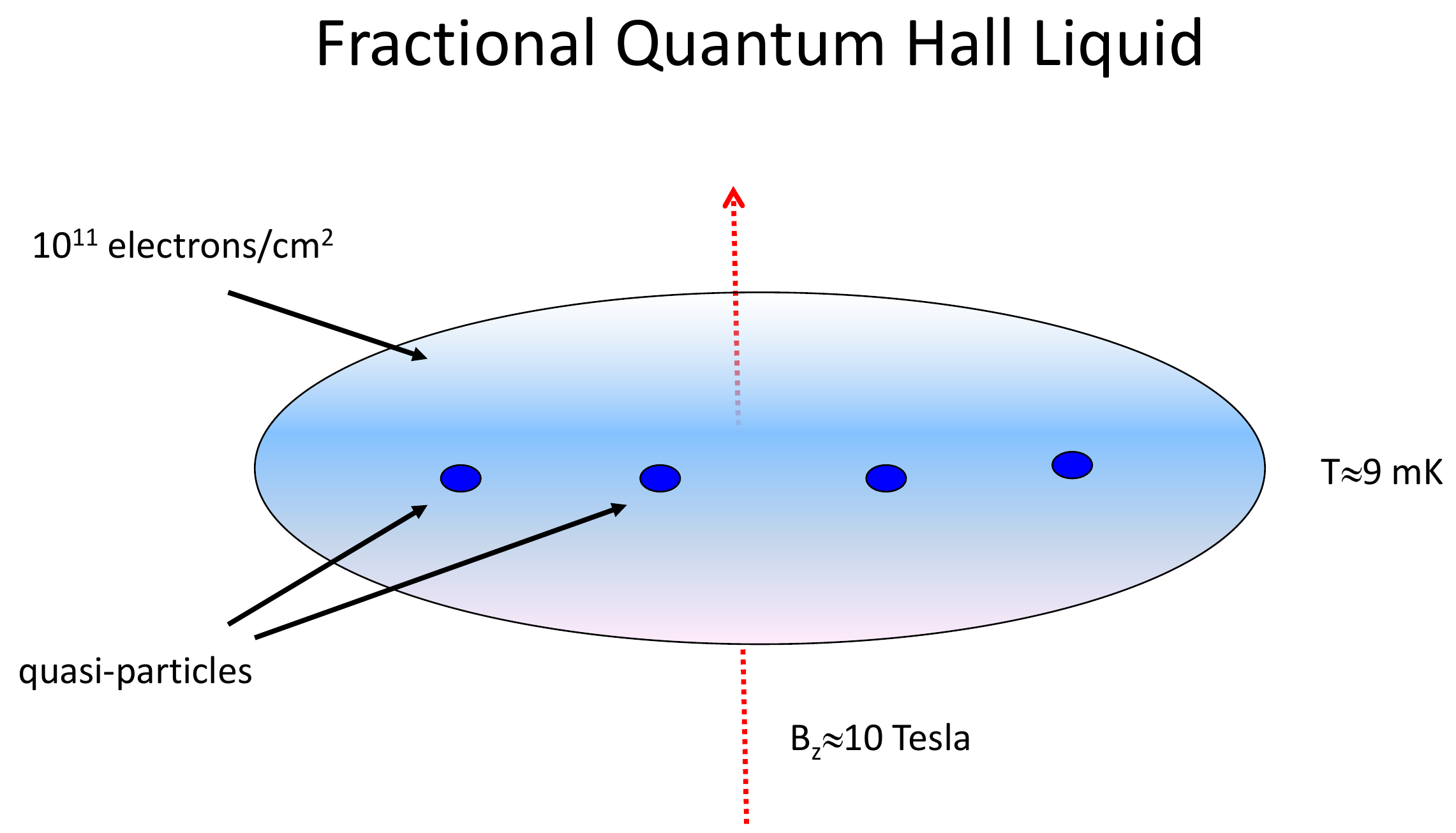}\hspace{2pc}
\begin{minipage}[b]{14pc} \caption{\label{fig1}Electrons are effectively confined to a 2D disk due to the low energy, perpendicular magnetic field causes quasi-particles to emerge.}\vspace{.5in}
\end{minipage}
\end{figure}

According to \cite{Nayaketal}: \textit{A system is in a topological
phase if, at low temperatures and energies and
long wavelengths, all observable properties e.g., correlation
functions are invariant under smooth deformations of the space-time manifold in
which the system lives,} or, alternatively, \textit{if its low energy effective field
theory is a topological quantum field theory TQFT,
i.e., a field theory whose correlation functions are invariant
under diffeomorphisms.}
The mathematics of anyons in 2D can be distilled down to the study of (quasi-)particles on surfaces.  How such systems may arise in nature is explained in \cite{Nayaketal}, we content ourselves to present Figure \ref{fig1} as a mathematician's cartoon of the fractional quantum Hall effect.  
The spin-statistics theorem of Fierz and Pauli shows that in 3 spacial dimensions, every particle is either a fermion or a boson: particle exchange of a system of two indistinguishable particles changes the state by at most a sign: $|\psi_1\psi_2\rangle=\pm|\psi_2\psi_1\rangle$.  Mathematically, this is related to the fact that the group of motions of two points in 3-dimensional space is $S_2$, the group of order $2$.  In 2 spacial dimensions the group of motions of $n$ points in a disk is the braid group $\B_n$, an infinite group with generators:
$ \sigma_1,\ldots,\sigma_{n-1}$ obeying:
\begin{enumerate}
 \item[(R1)]$\sigma_i\sigma_{i+1}\sigma_{i}=
\sigma_{i+1}\sigma_{i}\sigma_{i+1}$ for $1\leq i\leq n-2$ and
\item[(R2)] $\sigma_i\sigma_j=\sigma_j\sigma_i$
if $|i-j|>1$
\end{enumerate}
The suggests that \textit{exchange statistics of point-like particles in 2 spacial dimensions are richer than in 3 dimensions}.  Wilczek \cite{Wilczek} coined the term ``anyons'' to describe particles obeying exchange statistics $|\psi_1\psi_2\rangle=e^{2\pi\theta i}|\psi_2\psi_1\rangle$ for any $\theta$, where $\theta=0$ and $\theta=1/2$ correspond to bosons and fermions, respectively.  Since the time evolution of a closed system is unitary, particle exchange for $n$ indistinguishable anyons in the disk induces a unitary representation of the braid group $\B_n$, on the (Hilbert) state space of such configurations.  

\begin{figure}[h]
\includegraphics[width=3.6in]{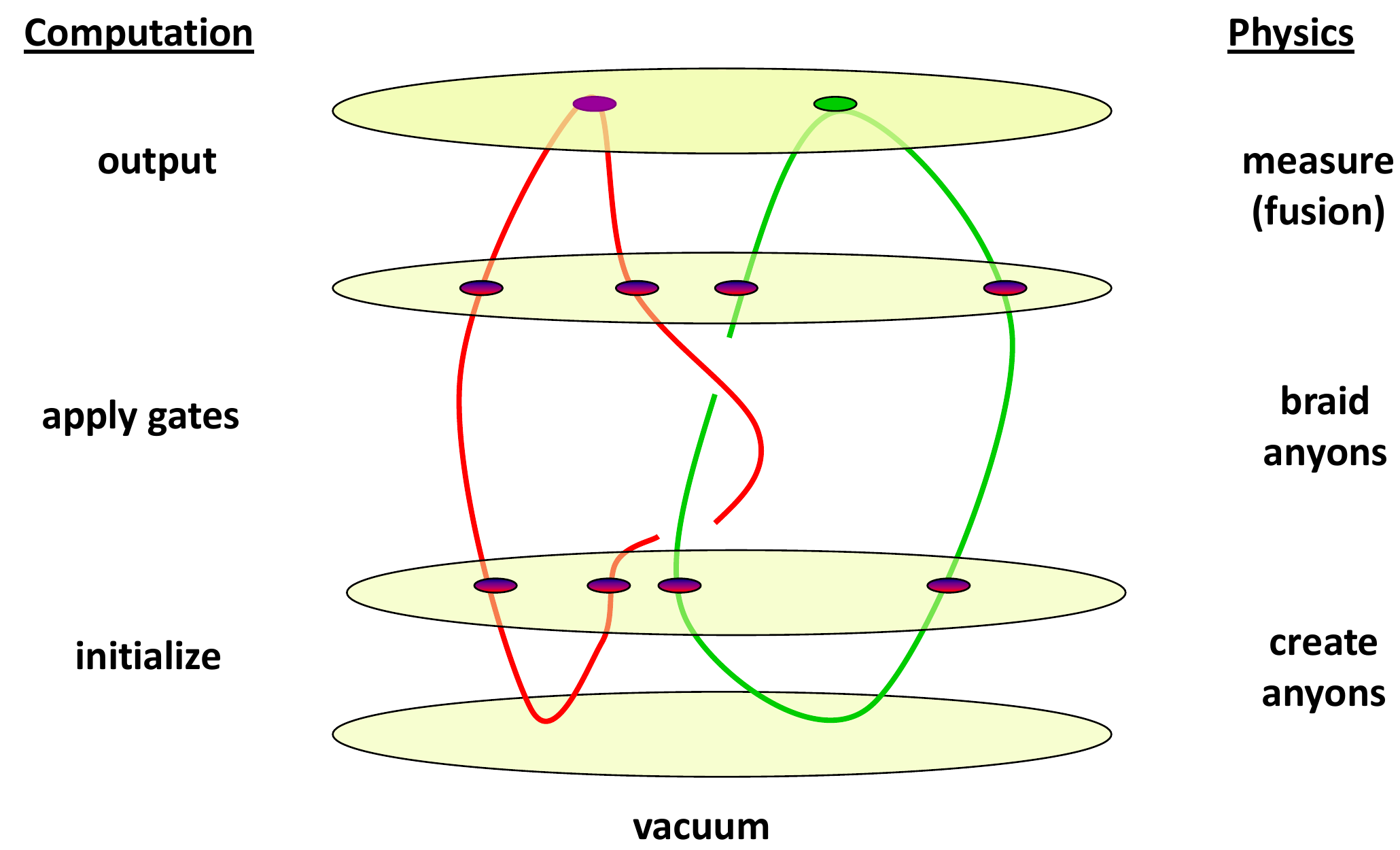}\hspace{2pc}
\begin{minipage}[b]{14pc} \caption{\label{fig2}A topological quantum computation scheme: particle-antiparticle pairs are drawn from the vacuum, particle exchange is performed followed by a measurement of the resulting particle type(s).}\vspace{0.5in}
\end{minipage}
\end{figure}

Assuming we can manufacture, manipulate and measure 2-dimensional topological phases of matter, we have a scheme for topological quantum computation (see Figure \ref{fig2}). 

Our first problem is now clear: how can we model a quantum mechanical system of anyons on a surface?  Two mathematical facts emerge from the quantum computational scheme in Figure \ref{fig2}: 1) to each surface with labeled points we must assign a Hilbert space in a consistent way (i.e. respecting the topological invariance and quantum mechanical constraints) and 2) particle exchange must correspond to a unitary operation on the associated Hilbert space.

\subsection{State spaces for anyons on surfaces}  
Let us explore the problem of assigning a Hilbert space $\cH(M,\ell)$ to a system of anyons $\ell$ on a surface $M$, in a way consistent with quantum mechanical principles and the topological nature of anyons.  What we will be describing is essentially the $2$-dimensional part of a $(2+1)$-TQFT, see \cite{Walker} for an excellent online resource.
Suppose our system supports exactly $r$ distinguishable indecomposable anyon types, which we label by a set $\LL=\{0,\ldots,r-1\}$ of ``colors'' or ``particle types''.  We include the vacuum as a sort of invisible anyon type which we label by $0$, or sometimes $\one$.  As we interpret the anyons as (quasi)-particles, each anyon $a$ must have an antiparticle, which we denote by $\hat{a}$.  Notice that the vacuum is its own antiparticle so $\hat{0}=0$.  
Anyons on a surface are point-like, so we may imagine them as small boundary circles on the surface labeled by $a\in\LL$.  A convenient topological interpretation of anti-particles is that a positively oriented circle labeled by $a$ is the same as a negatively oriented circle labeled by $\hat{a}$.  Topological invariance means that any topologically allowed operations such as anyon exchange, or twisting part of a surface must correspond to a unitary operator on the corresponding Hilbert space.  The underlying Hilbert space must remain the same, provided the operation leaves the topology of the surface $M$ with labeled punctures the same.

\begin{figure}[h]
\includegraphics[width=1.6in]{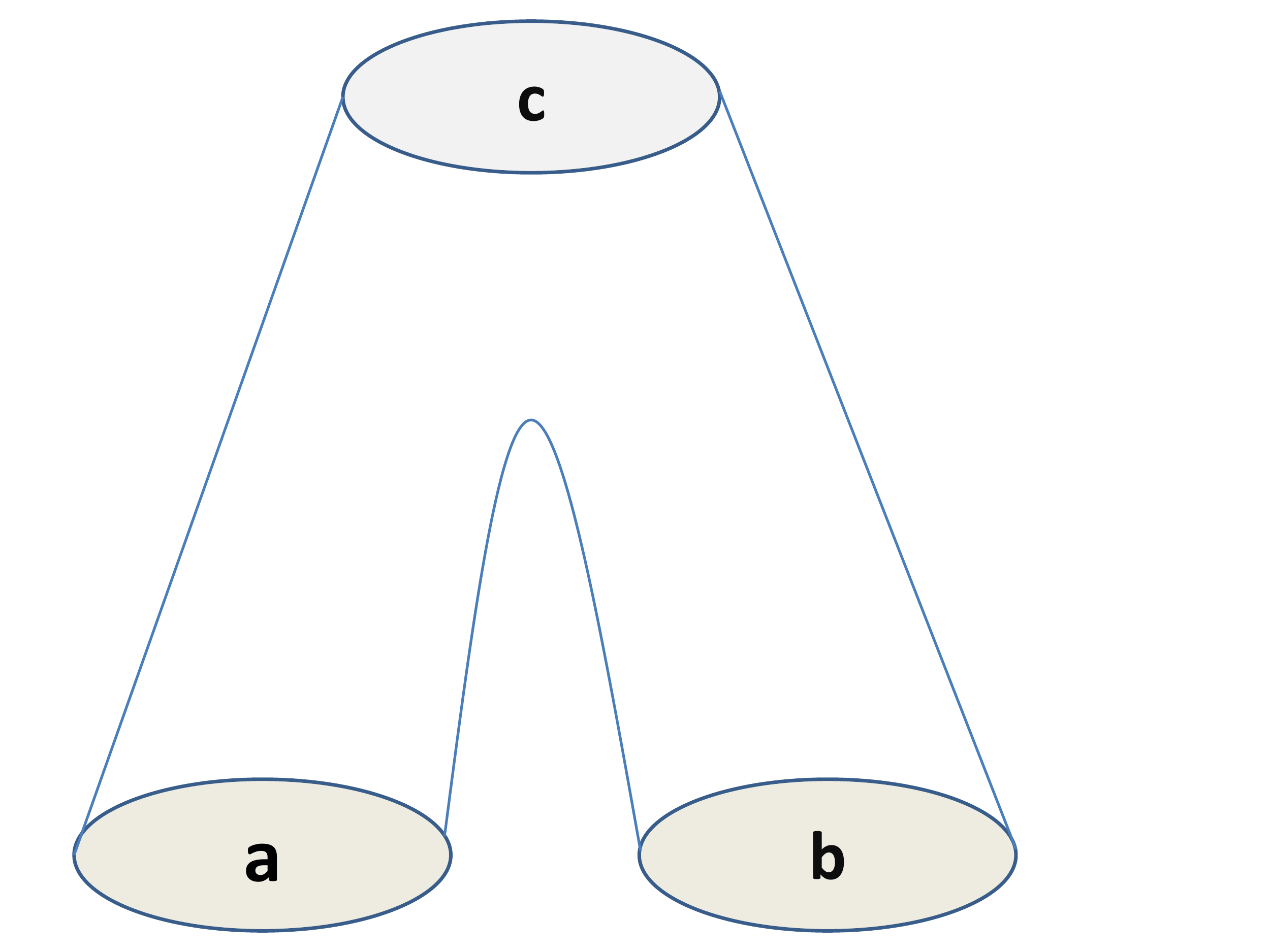}\hspace{2pc}
\begin{minipage}[b]{16pc} \caption{\label{fig3} The labeled surface associated with the fusion channel $H_{ab}^c$, which is to be read upwards, and interpreted as the number of independent ways that particles $a$ and $b$ can fuse to give particle $c$.}\vspace{0.2in}
\end{minipage}
\end{figure}

The notions of \textit{entanglement}, (particle-anti-particle) \textit{duality} and  \textit{locality} lead us to the following:
\begin{itemize}
\item \textbf{Disjoint Union Axiom} The Hilbert space of two disjoint systems (surfaces with labeled boundary) $(M_1,\ell_1)$ and $(M_2,\ell_2)$, considered as a composite system is $\cH(M_1,\ell_1)\otimes\cH(M_2,\ell_2)$.
\item \textbf{Duality Axiom} $\cH(M,\ell)^\dag\cong \cH(\overline{M},\hat{\ell})$, where $\overline{M}$ is the surface $M$ with opposite orientation and $\hat{\ell}$ means apply $\hat{}$ to each $x\in\ell$.
 \item \textbf{Gluing Axiom} The global Hilbert space is determined by local Hilbert spaces.  More precisely, if a surface $(M,\ell)$ with boundary labels $\ell$ is obtained from $(M_g,x,\hat{x},\ell)$ by gluing the boundary circles labeled by $x$ and $\hat{x}$ together, then $$\cH(M,\ell)=\bigoplus_{x\in\LL}\cH(M_g,x,\hat{x},\ell).$$
\end{itemize}
Taken together, these axioms point us to a cut-and-paste procedure to determine the Hilbert space associated to a surface with labeled boundary from ``initial conditions'', i.e. the Hilbert spaces of a few less complicated surfaces.  The first three initial conditions are:
\begin{itemize}
 \item \textbf{Empty Set Axiom} $\cH(\emptyset)\cong \C$
 \item \textbf{Disk Axiom} $\cH(D^2,x)\cong \delta_{0x}\C$
 \item \textbf{Annulus Axiom} $\cH(A,x,y)\cong\delta_{x\hat{y}}\C$
\end{itemize}
As axioms these are meant to be unquestioned--however they can be justified as consistency.  For example, the Disk and Annulus Axioms can be understood as particle-anti-particle creation rules.  The last initial condition(s) are meant to determine the Hilbert spaces for all surfaces with labeled boundary, subject to compatibility with the previous axioms:
\begin{itemize}
 \item \textbf{Pants Axiom} $\cH(P(z_1,z_2,z_3),x,y,z)=\C^{N_{xy}^z}$ for some $N_{x,y}^z\geq 0$.
\end{itemize}
Here $P(z_1,z_2,z_3):=S^2\smallsetminus\{z_1,z_2,z_3\}$ is the sphere with three boundary components (``a pair of pants'') labeled by $x,y$ and $z$.  The problem now is to choose the $N_{xy}^z$ for all triples of labels $(x,y,z)\subset \LL^3$ compatible with all of the axioms.  The alert reader will realize that the pants axiom does not uniquely assign a Hilbert space to labeled pair of pants.  Indeed, there is a $6$-fold ambiguity: one for each permutation of the labels $x,y$ and $z$!  This technicality is dealt with by replacing surfaces $M$ by so-called \textit{$m$-surfaces} or \textit{extended surface} which have a bit more structure. The idea is that one should fix a particular ``standard'' surface with boundary and use it to parametrize all surfaces with boundary that are homeomorphic the the standard surface.  The parametrization then removes the ambiguity.  This is carefully addressed in \cite[Chapter 4]{TuraevBook} or \cite[Chapter 4]{BK}.  For a sphere with punctures it amounts to choosing some 
labels as ``inputs'' and others as ``outputs'' and carefully ordering the labels along two parallel lines.  For two input labels $a$ and $b$ and one output label $c$ these give us \textit{fusion channels} $H_{ab}^c$: a Hilbert space of dimension $N_{ab}^c$, see Figure \ref{fig3}.  

The interpretation of the Hilbert space associated with the surface in Figure \ref{fig3} as a fusion channel suits our purposes very well: we would like to imagine these as processes.
Dual to fusion channels are the \textit{splitting channels} with $1$ input (say $c$) and $2$ outputs (say $a$ and $b$)  denoted $H_c^{ab}$.  Using the axioms above, \textit{every} state space can be obtained via tensor products and direct sums of splitting and fusion channels. Which fusion/splitting channels correspond to the annulus axiom?  By choosing the fusion output or splitting input to be the vacuum $0$ label we get $H_{ab}^0\cong\delta_{b,\hat{a}}\C\cong H_{0}^{ab}$.  
\begin{exercise}
 Use the gluing axiom and the annulus axiom to show that the state space with input and output $a$ has: $H_{a0}^a\cong\C\cong H_a^{a0}$.
\end{exercise}
The precise sense in which these space are dual will be described below.

In general $N_{xy}^z\neq N_{xz}^y$: this is due to the technicality mentioned above.  However, using the annulus, gluing and disjoint union axioms one can show that $N_{x\hat{z}}^{\hat{y}}=N_{xy}^z$: one simply glues cylinders (annuli) with labels $z$ and $y$ to change them from inputs to outputs and vice versa.  Similarly the duality axiom implies that $N_{xy}^z=N_{\hat{x}\hat{y}}^{\hat{z}}$.  In fact, if we assume that every particle is indistinguishable from its anti-particle (i.e. $\hat{x}=x$), the \textit{dimension} of $\cH(M,\ell)$ can be computed unambiguously: for then the $N_{xy}^z$ are fully symmetric in the three indices.

To get a better intuition for how these axioms allow one to determine (at least the dimension of) the Hilbert space of any surface with labeled boundary we suggest trying the following:
\begin{exercise}
 Show that $\dim\cH(T^2)=|\LL|$, where $T^2$ is the 2-dimensional torus.
\end{exercise}

\begin{figure}[h]
\includegraphics[width=2.7in]{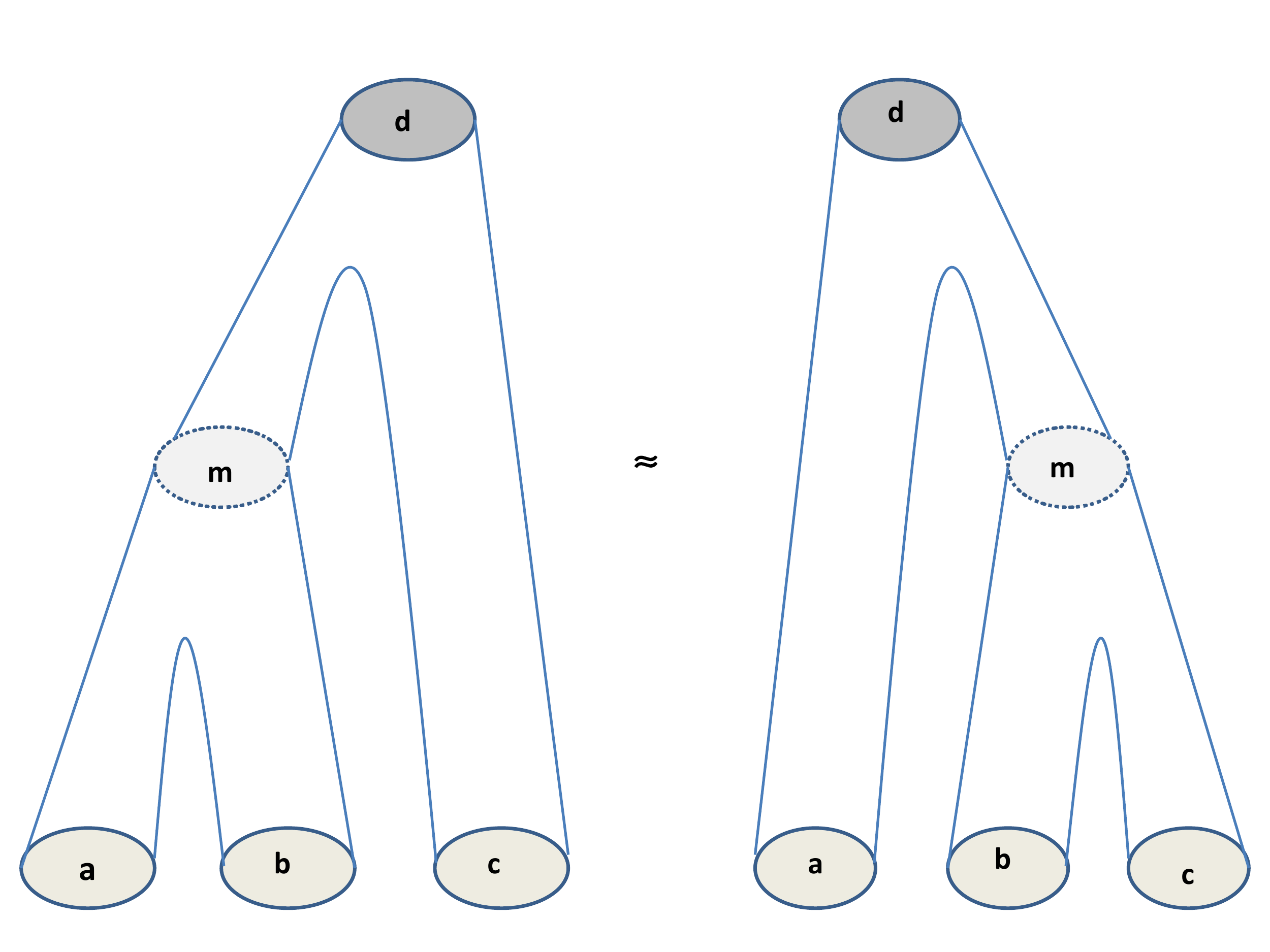}\hspace{2pc}
\begin{minipage}[b]{18pc} \caption{\label{fig4} Equivalent (extended) labeled surfaces.  The dimension of each can be computed via the gluing (over $m$) and disjoint union axioms, and then interpreted as a matrix product.  For example the the LHS is: $\sum_m N_{ab}^mN_{\hat{m}c}^d$. As a consequence one finds $N_aN_c=N_cN_a$.}\vspace{0.4in}
\end{minipage}
\end{figure}

Topological invariance provides many useful constraints on the numbers $N_{xy}^z$, since these are interpreted as dimensions of Hilbert spaces associated with a thrice punctured sphere. For example, interchanging labels $x$ and $y$ (i.e. braiding, but keeping both $x$ and $y$ as inputs) is a topological (commutativity) operation, so $N_{xy}^z=N_{yx}^z$.  To give algebraic interpretations we define, for each label $a$, a \textit{fusion} matrix $N_a$ by $(N_a)_{c,b}=N_{ab}^c$.  In this formalism the above calculations imply that $N_{\hat{x}}=N_x^\dag$.  The \textit{fusion rules} express the fusion channels as a superposition of particle types that can occur as an output with inputs $x$ and $y$: $x\times y=\sum_z N_{xy}^z z$.  A cascade of fusion channels can then be interpreted as a matrix product.   A topological associativity constraint and a consequence for matrices is illustrated in Figure \ref{fig4}.

\subsection{Initialization, transformations and measurement} 
\begin{figure}[h]\includegraphics[width=2.7in]{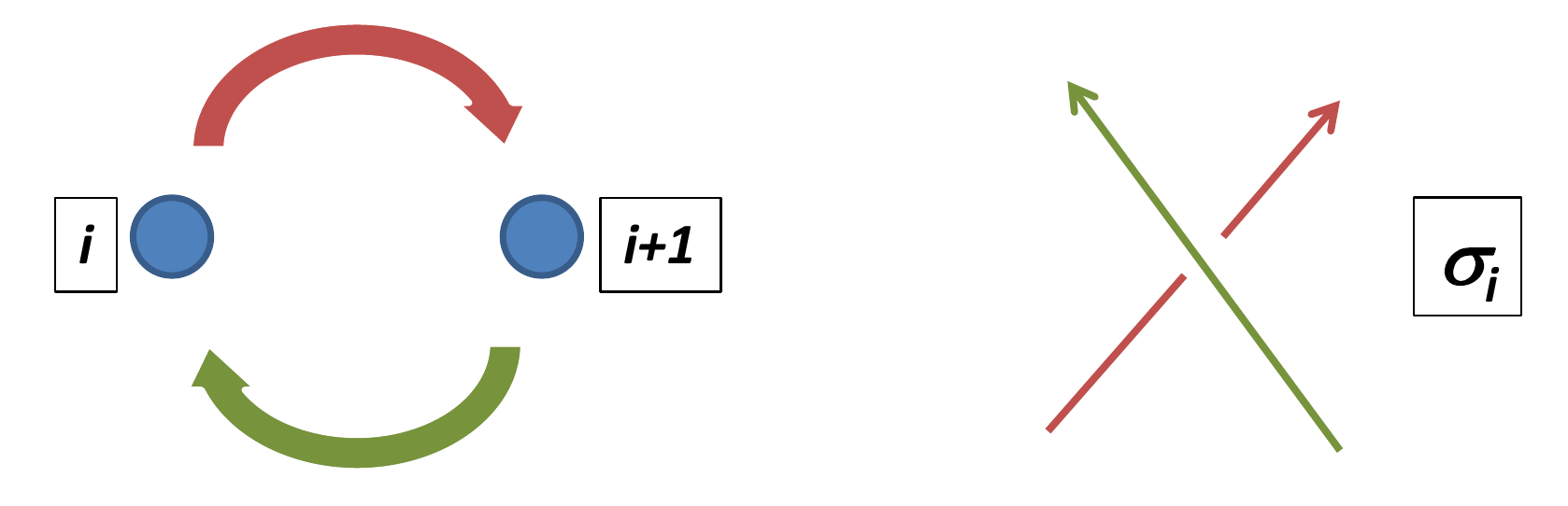}\hspace{2pc}
\begin{minipage}[b]{20pc} \caption{\label{fig5} Interchanging the positions of two identical particles induces a quantum gate--the image of $\sigma_i$ under a unitary representation on the state space. }\vspace{.1in}
\end{minipage}
\end{figure}

So far we have only discussed the spaces of states.  
We can now proceed to explore the unitary operations (quantum gates), processes and further constraints in our model.  To do this, we need to establish notation for states themselves.
How do we describe a specific state in, for example, the fusion channel $H_{ab}^c$ as in Figure \ref{fig3}?   It is convenient to denote such a vector by the skeleton of the corresponding space, i.e. the trivalent graph {\large \rotatebox[origin=c]{180}{$\mathsf{Y}$}} with the three extremal vertices labeled by $a,b$ and $c$ and the degree three vertex labeled to distinguish it from other states in $H_{ab}^c$.  For example, we might choose a basis for $H_{ab}^c$, so that there are $\dim(H_{ab}^c)$ labels.  Similarly, we use appropriately labeled graphs {\large $\mathsf{Y}$} to denote states in the dual space $H_{c}^{ab}$.  It is tempting, and indeed can be justified mathematically using the gluing axiom, to stack these graphs to represent a cascade of splitting/fusion operations.  For example, if we compose compatibly labeled {\large $\mathsf{Y}$} (input $c$, outputs $a,b$) and {\large \rotatebox[origin=c]{180}{$\mathsf{Y}$}} (inputs $a,b$, output $c$) the result is a vector in the $1$-dimensional state 
space $H_{c0}^c\cong\C$.  One typically choosing the bases so that this pairing coincides with the inner product on the Hilbert space $H_{ab}^c$.  A complete treatment of this diagrammatic yoga of \textit{graphical calculus} involving such pictures can be found in \cite[Appendix E]{Kitaev} and \cite[Section 4.2]{WangBook}. See also the discussion of the topological twist below and Figure \ref{fig5} for the picture associated with the braiding operators.  Some calculations of this form are illustrated in Figures \ref{fig8} and \ref{fig9}.

 The vacuum state corresponds to a disk with boundary labeled by $0$, which we can use as an invisible input or output without changing the state.  The creation of a particle-antiparticle from the vacuum corresponds to a disk with boundary labeled by $0$ and two interior boundary circles labeled by $a$ and $\hat{a}$.  This process translates to a linear (Hermitian) operator on state spaces: $b_a:H_{00}^0\rightarrow H_{0}^{a\hat{a}}$, with a corresponding (dual) annihilation $d_a:H_{0}^{a\hat{a}}\rightarrow H_{00}^0$.  Indeed, we assume we can create (via some physical process) any number of particle-antiparticle pairs, from which we obtain, from the vacuum state, a state in $H_0^{a_1\hat{a}_1\cdots a_n\hat{a}_n}$.  This corresponds to \textit{initialization} in the quantum computational model.  On the other hand, if we are given a system with state vector in $H^{b_1\cdots b_m}_{a_1\cdots a_n}$ we assume 
we can measure the total charge (i.e. the label) of a pair of adjacent particles, perhaps by bringing them together and measuring the energy.  This is the \textit{measurement} stage of the quantum computation, a Hermitian operator represented graphically as composing with a fusion operator 
{\large $\mathsf{Y}$}.

The time evolution of the space of states must be a unitary operator.  In particular, a sequence of particle exchanges corresponding to a braid $\beta$ induces a unitary transformation 
$|\psi\rangle\mapsto U_\beta|\psi\rangle$.  In the topological model of \cite{FKLW} these are (all of)\footnote{Recently some models employing partial measurement \cite{CuiWang} have been explored, but for the sake of simplicity we will only consider braiding operators as our quantum circuits.} the \textit{quantum circuits}.   If we consider a simple case where $a=\hat{a}$, then the state space $H_0^{a,\ldots,a}$ corresponding to $n$ particles of type $a$ supports a unitary representation of the braid group $\B_{n}$ via particle exchange, as illustrated in Figure \ref{fig5}. 
More generally, we always obtain a unitary representation of the small group of \textit{pure braids} consisting of those braids with each strand beginning and ending in the same position.  The computational strength of the model is hidden in this unitary representation of $\B_n$.  A particle $a$ is called \textbf{non-abelian} if the image of the $\B_n$ representation on the state space of $n$ type $a$ particles is non-abelian (for some $n$).  To have a reasonable computational model this is a bare minimum.  An anyon $a$ is called \textbf{(braiding) universal} if any unitary operator can be approximately achieved as the image of some braid $\beta$ acting on a state space of $n$ type $a$ particles via particle exchange (plus some technical ``no-leakage'' condition that we ignore).  The search for non-abelian and universal anyons is a major thrust of experimental condensed matter physics.

To summarize the processes we assume are available: 1) we can create any number of particle-antiparticle pairs, 2) we may exchange these particles to rotate our initial state and 3) we may measure the particle type of any pair of neighboring particles.  One key is that after braiding the particles' world lines, a neighboring particle-antiparticle pair may have obtained a different total charge (besides $0$, i.e. the vacuum).  To get meaningful information from this process we must repeat the same process several times, taking a tally of the outputs (particle types).  The topological degrees of freedom ensure that slight variations in the process (e.g. small deviations in the trajectory of a particle in space-time) do not influence the output.  The empirically computed probability distribution of output particle types constitutes the result of the quantum computation.

\section{Fundamental Questions}
In the remainder of this survey we would like to address a few fundamental questions:
\begin{enumerate}
 \item How can we distinguish indecomposable particle types?
 \item Is there a ``periodic table'' of topological phases of matter?
 \item How can we detect non-abelian and universal anyons in (idealized) experiments?

\end{enumerate}

\subsubsection{Distinguishing particles}
For any two particles types $a$ and $b$, we must have an (idealized) quantum process that distinguishes the particle types.  Essentially, we need to be able to determine some unknown particle type using creation, braiding and measurement.  \begin{figure}[h]\includegraphics[width=2.6in]{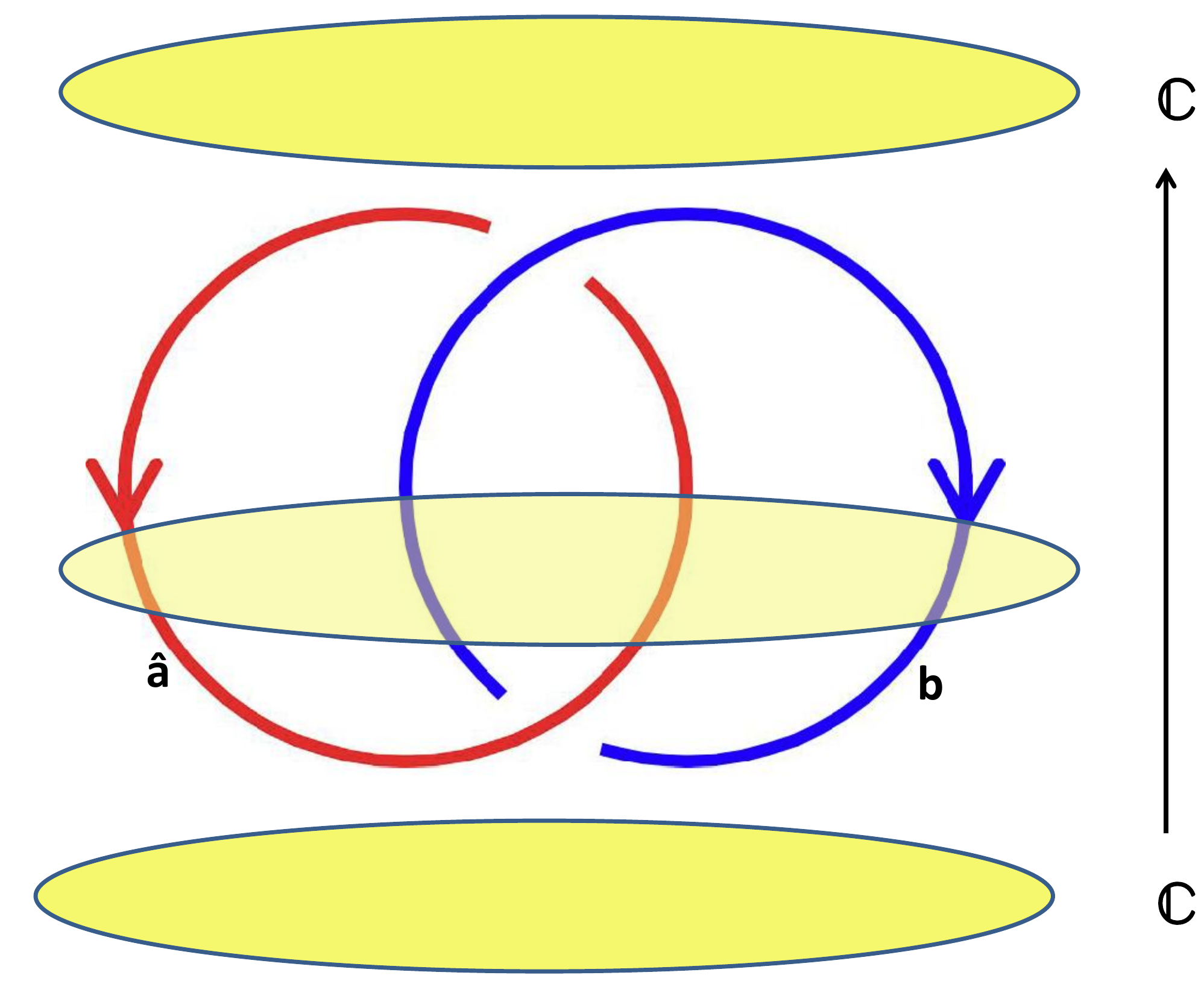}\hspace{2pc}
\begin{minipage}[b]{20pc} \caption{\label{fig6}Creating particle-antiparticle pairs of types $a$ and $b$, braiding and then measuring the amplitude of the vacuum output produces a constant map $S_{ab}\in\C$ for each $a,b\in\LL$.  For fixed $a$, we require that the vector of outputs over all $b\in\LL$ be linearly independent with (in fact, orthogonal to) that of any other $a^\prime$.  That is, the matrix $S$ must be orthogonal (up to an overall phase).}\vspace{.3in}
\end{minipage}
\end{figure}Figure \ref{fig6} illustrates the process, which leads to a certain non-degeneracy constraint on the braiding.  The columns of the $S$-matrix can be seen to be simultaneous eigenvectors for the (commuting) fusion matrices $\{N_a:a\in\LL\}$, so that $S$ diagonalizes the $N_a$.  This leads to the famous \textit{Verlinde formula}: $N_{ij}^k=\frac{1}{D^2}\sum_r\frac{S_{ir}S_{jr}S_{hat{k}r}}{S_{0r}}$ where $D^2$ is an overall normalization constant.

Anyons may sometimes also be distinguished by their \textit{topological spin}, the phase acquired on the (1-dimensional) state space of the cylinder labeled by $a$ upon twisting by $2\pi$.  For this one should imagine that the particle $a$ is a small line segment so that the world lines are really ribbons rather than 1-dimensional curves.  Then if we twist $a$ by $2\pi$ radians, its trajectory traces out a narrow ribbon with a twist.  Depicting trajectories as curves we have \[ \xy
(0,10)*{}="T"; (0,-10)*{}="B"; (0,5)*{}="T'"; (0,-5)*{}="B'";
"T";"T'" **\dir{-}; "B";"B'" **\dir{-}; (7,0)*{}="LB"; "B'";"LB"
**\crv{(1,4) & (7,4)}; \POS?(.25)*{\hole}="2z"; "LB"; "2z"
**\crv{(8,-6)*{a} & (2,-6)}; "2z"; "T'" **\crv{(0,-3)}; (12,0)*{}="S";
(14,0)*{}="S'"; "S";"S'" **\dir2{-}; (18,0)*{\theta_a};
(22,10)*{}="U"; (22,-10)*{a}="V"; "U";"V" **\dir{-};
\endxy \]
where $\theta_a=e^{2\pi ih_a}$ and $h_a\in\Q$ is the topological twist.  We must take care to remember that the picture on the left acquires a phase $\theta_a$ when pulled straight.

\subsubsection{Periodic table}
How many distinct models of anyonic systems with exactly $n=|\LL|$ distinct particle types are there?  More generally, is there a classification of such models?  Using extensive algebraic constraints we \cite{BNRW1} recently proved that, for fixed $n$, there are finitely many possible models.  The precise asymptotics of the number of distinct theories as $n\rightarrow\infty$ are unknown, but it can be shown that it grows faster than any polynomial.
A classification up to $|\LL|\leq 5$ is known (see \cite{BNRW2} and references) with constructions coming from quantum groups and finite groups, see Table \ref{table1}.
 \begin{table}[h]\vspace{-.15in}
\begin{tabular}{|c|c|}
\hline
$|\LL|$ & Models\\
\hline\hline $1$ & Vec\\\hline $2$ & $Fib$, $\Z_2$ \\\hline
$3$ & $\Z_3,PSU(2)_7$, Ising \\
\hline
$4$ & products, $\Z_4$, $PSU(2)_9$\\\hline
$5$ & $\Z_5,PSU(2)_{11},SU(3)_4/\Z_3,SU(2)_4$
\\
\hline 
\end{tabular}\hspace{2pc}
\begin{minipage}[b]{18pc} \caption{\label{table1}$SU(N)_k$ are the level $k$ representations of the affine Kac-Moody algebra of type $A_{N-1}$, and $PSU(2)_k$ consists of the ``integer spin'' representations.  The $\Z_n$ models are abelian--each fusion channel is 1-dimensional, with fusion rules like the multiplication in $\Z_n$. 
}\vspace{-.5in}
\end{minipage}
\end{table}
The following are two explicit examples:
\begin{example}  The \textbf{Fibonacci} theory has two labels
 $\LL=\{\one,f\}$ with fusion rules $f\times f=\one+f$.  The $S$-matrix and topological twists are: $S=\begin{pmatrix} 1 & \frac{1+\sqrt{5}}{2}\\\frac{1+\sqrt{5}}{2} & -1\end{pmatrix}$ and $\theta_f=e^{4\pi i/5}$.  The name comes from the fact that $f^n=F_{n-1}\one+F_{n}f$ where $F_i$ is the well-known Fibonacci sequence: $0,1,1,\ldots$.
\end{example}
\begin{example} The \textbf{Ising} theory has three labels
 $\LL=\{\one,\sigma,\psi\}$ and fusion rules $\sigma\times\sigma=\one+\psi$, $\sigma\times\psi=\sigma$, $\psi\times\psi=\one$. The $S$-matrix and twists are $S=\begin{pmatrix} 1 & \sqrt{2} & 1\\ \sqrt{2} &0&-\sqrt{2}\\ 1&-\sqrt{2} & 1\end{pmatrix}$ and $\theta_\sigma=e^{\pi i/8}$, $\theta_\psi=-1$.  The $\psi$ particle is the famous Majorana fermion.
\end{example}
\subsubsection{Detecting Non-abelian and universal anyons}
\begin{figure}[h]
\begin{tikzpicture}[scale=0.33]
 \begin{scope}
 \begin{scope}[yshift=0cm]
    \draw [thick,dashed](0,-2)--(0,0);
    \draw (0,0)--(1,1);
    \draw (0,0)--(-1,1);
    \draw (0,-2) node[anchor=east] {$\mathbf{1}$};
    \draw (-1,1) node[anchor=north east] {${a}$};
    \draw (1,1) node[anchor=north west] {${\hat{a}}$};
  \end{scope}
  \begin{scope}[xshift=6cm]
    \draw  (0,-2)--(0,0);
    \draw (0,0)--(1,1);
    \draw (0,0)--(-1,1);
    \draw (0,-2) node[anchor=west] {${b}$};
    \draw (-1,1) node[anchor=north east] {${a}$};
    \draw (1,1) node[anchor=north west] {${\hat{a}}$};
  \end{scope}
  \begin{scope}[xshift=-1cm,yshift=1cm]
    \draw (8,0)--(2,6);
    \draw (6,0)--(0,6);
    \draw [white, line width=3mm] (0,0)--(6,6);
    \draw (0,0)--(6,6);
    \draw [white, line width=3mm] (2,0)--(8,6);
    \draw (2,0)--(8,6);
  \end{scope}
  \begin{scope}[yshift=8cm]
    \draw (0,0)--(0,2);
    \draw (-1,-1)--(0,0);
    \draw (1,-1)--(0,0);
    \draw (0,2) node[anchor=east] {${b}$};
    \draw (-1,-1) node[anchor=south east] {${a}$};
    \draw (1,-1) node[anchor=south west] {${\hat{a}}$};
  \end{scope}
  \begin{scope}[xshift=6cm, yshift=8cm]
    \draw [thick,dashed] (0,0)--(0,2);
    \draw (-1,-1)--(0,0);
    \draw (1,-1)--(0,0);
    \draw (0,2) node[anchor=west] {$\mathbf{1}$};
    \draw (-1,-1) node[anchor=south east] {${a}$};
    \draw (1,-1) node[anchor=south west] {${\hat{a}}$};
  \end{scope}
 \end{scope}
\begin{scope}[xshift=11cm,yshift=4cm]
 \draw (0,0) node {$=~~\alpha$};
\end{scope}
\begin{scope}[xshift=16cm, yshift=1cm]
  \begin{scope}[yshift=0cm]
   \draw [thick] (0,-1.5)--(0,0);
   \draw (0,0)--(1.5,1.5);
   \draw (0,0)--(-1.5,1.5);
   \draw (0,-1.5) node[anchor=east] {${b}$};
   \draw (-1.5,1.5) node[anchor=east] {${a}$};
   \draw (1.5,1.5) node[anchor=west] {${\hat{a}}$};
  \end{scope}
  \begin{scope}[yshift=3cm]
   \draw (-1.5,-1.5)--(-1.5,1.5);
   \draw (1.5, -1.5)--(1.5,1.5);
  \end{scope}
 \begin{scope}[yshift=6cm]
  \draw (0,0)--(0,1.5);
  \draw (-1.5,-1.5)--(0,0);
  \draw (1.5,-1.5)--(0,0);
  \draw (0,1.5) node[anchor=east] {${b}$};
\end{scope}
\end{scope}
\begin{scope}[xshift=22cm,yshift=4cm]
 \draw (0,0) node {$\neq~~0$};
\end{scope}
\end{tikzpicture}\hspace{2pc}
\begin{minipage}[b]{13pc} \caption{\label{fig8} The top loop may be deleted at the expense of a non-zero scalar $\alpha$, yielding the non-zero state on the right.}\vspace{.6in}
\end{minipage}
\end{figure}
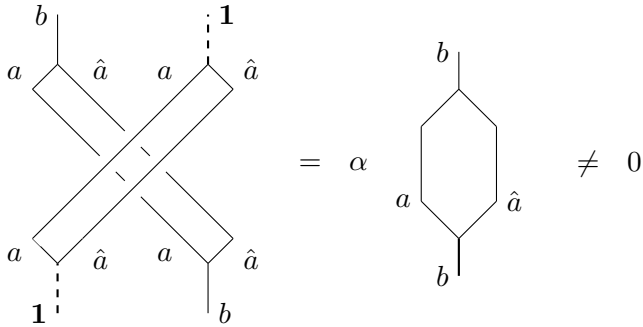

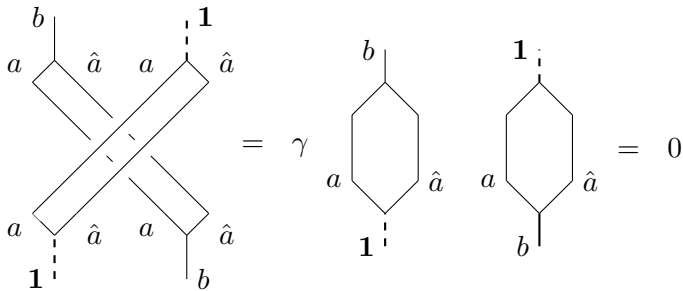
\begin{figure}[h]

\begin{tikzpicture}[scale=0.29]
 \begin{scope}
  \begin{scope}[yshift=0cm]
    \draw [thick,dashed](0,-2)--(0,0);
    \draw (0,0)--(1,1);
    \draw (0,0)--(-1,1);
    \draw (0,-2) node[anchor=east] {$\mathbf{1}$};
    \draw (-1,1) node[anchor=north east] {${a}$};
    \draw (1,1) node[anchor=north west] {${\hat{a}}$};
  \end{scope}
  \begin{scope}[xshift=6cm]
    \draw  (0,-2)--(0,0);
    \draw (0,0)--(1,1);
    \draw (0,0)--(-1,1);
    \draw (0,-2) node[anchor=west] {${b}$};
    \draw (-1,1) node[anchor=north east] {${a}$};
    \draw (1,1) node[anchor=north west] {${\hat{a}}$};
  \end{scope}
  \begin{scope}[xshift=-1cm,yshift=1cm]
    \draw (8,0)--(2,6);
    \draw (6,0)--(0,6);
    \draw [white, line width=3mm] (0,0)--(6,6);
    \draw (0,0)--(6,6);
    \draw [white, line width=3mm] (2,0)--(8,6);
    \draw (2,0)--(8,6);
  \end{scope}
  \begin{scope}[yshift=8cm]
    \draw (0,0)--(0,2);
    \draw (-1,-1)--(0,0);
    \draw (1,-1)--(0,0);
    \draw (0,2) node[anchor=east] {${b}$};
    \draw (-1,-1) node[anchor=south east] {${a}$};
    \draw (1,-1) node[anchor=south west] {${\hat{a}}$};
  \end{scope}
  \begin{scope}[xshift=6cm, yshift=8cm]
    \draw [thick,dashed] (0,0)--(0,2);
    \draw (-1,-1)--(0,0);
    \draw (1,-1)--(0,0);
    \draw (0,2) node[anchor=west] {$\mathbf{1}$};
    \draw (-1,-1) node[anchor=south east] {${a}$};
    \draw (1,-1) node[anchor=south west] {${\hat{a}}$};
  \end{scope}
 \end{scope}
\begin{scope}[xshift=10cm,yshift=4cm]
 \draw (0,0) node {$=~~\gamma$};
\end{scope}
\begin{scope}[xshift=15cm, yshift=1cm]
  \begin{scope}[yshift=0cm]
   \draw [thick, dashed] (0,-1.5)--(0,0);
   \draw (0,0)--(1.5,1.5);
   \draw (0,0)--(-1.5,1.5);
   \draw (0,-1.5) node[anchor=east] {$\mathbf{1}$};
   \draw (-1.5,1.5) node[anchor=east] {${a}$};
   \draw (1.5,1.5) node[anchor=west] {${\hat{a}}$};
  \end{scope}
  \begin{scope}[yshift=3cm]
   \draw (-1.5,-1.5)--(-1.5,1.5);
   \draw (1.5, -1.5)--(1.5,1.5);
  \end{scope}
 \begin{scope}[yshift=6cm]
  \draw (0,0)--(0,1.5);
  \draw (-1.5,-1.5)--(0,0);
  \draw (1.5,-1.5)--(0,0);
  \draw (0,1.5) node[anchor=east] {${b}$};
\end{scope}
\end{scope}
\begin{scope}[xshift=22cm, yshift=1cm]
  \begin{scope}[yshift=0cm]
   \draw [thick] (0,-1.5)--(0,0);
   \draw (0,0)--(1.5,1.5);
   \draw (0,0)--(-1.5,1.5);
   \draw (0,-1.5) node[anchor=east] {${b}$};
   \draw (-1.5,1.5) node[anchor=east] {${a}$};
   \draw (1.5,1.5) node[anchor=west] {${\hat{a}}$};
  \end{scope}
  \begin{scope}[yshift=3cm]
   \draw (-1.5,-1.5)--(-1.5,1.5);
   \draw (1.5, -1.5)--(1.5,1.5);
  \end{scope}
 \begin{scope}[yshift=6cm]
  \draw [thick, dashed](0,0)--(0,1.5);
  \draw (-1.5,-1.5)--(0,0);
  \draw (1.5,-1.5)--(0,0);
  \draw (0,1.5) node[anchor=east] {$\mathbf{1}$};
\end{scope}
\end{scope}
\begin{scope}[xshift=27cm,yshift=4cm]
 \draw (0,0) node {$=~~0$};
\end{scope}
\end{tikzpicture}\hspace{2pc}
\begin{minipage}[b]{13pc} \caption{\label{fig9} If the exchange of a pair of $a$ (respectively $\hat{a}$) particles and the full position interchange of a $a$-$\hat{a}$ pair are each multiples of the identity operation we produce a zero state by the annulus axiom.}\vspace{.3in}
\end{minipage}
\end{figure}

The \textbf{quantum dimension} $\dim(a)$ of a particle type $a$ is the maximal eigenvalue of the fusion matrix $N_a$.  By the Perron-Frobenius theorem in matrix theory, this eigenvalue is real and positive.  In fact, it can be shown that $\dim(a)\geq 1$, since no power of $N_a$ is $0$.    The Fibonacci particle $f$ has $\dim(f)=\frac{1+\sqrt{5}}{2}$ whereas the Ising particle $\sigma$ has $\dim(\sigma)=\sqrt{2}$.  If $\dim(a)>1$ we say that $a$ is \textbf{non-degenerate}: in this case $a\times \hat{a}=\one+b$ where $b\neq \one$.  Recently we \cite{nab} showed that non-degeneracy of $a$ implies $a$ is non-abelian.  The essence of the argument is illustrated in Figures \ref{fig8} and \ref{fig9}.  If we suppose that the braiding operators commute then we may simultaneously diagonalize them.  Restricting to each irreducible sector we may further assume that they act as scalar multiples of the identity.

Thus it is impossible that the braiding operators commute, since this contradicts the calculation in Figure \ref{fig8} of a non-zero state.  This result show that, in principle, experimentalists can detect non-abelian anyons by measuring the quantum dimension.  

It is known \cite{FLW2} that the Fibonacci anyon is universal, whereas the Ising anyon is not, despite the fact that both are non-degenerate (and hence non-abelian).  The braid group image corresponding to an array of Ising anyons is non-abelian, but finite.  How can we distinguish these models?  Over the last few years we have found significant evidence for the following:
\begin{conj}
 The anyon $a$ is (braiding) universal if, and only if, $\dim(a)^2$ is not an integer.
\end{conj}
One strong piece of evidence for this conjecture is that, for models associated with quantum groups, the braid group image is infinite if and only if $\dim(a)^2$ is not an integer.  This latter weaker version of the conjecture goes by the name \textit{property \textbf{F}} (see \cite{NR}).

\section{Three-dimensional generalizations}
\begin{center}\begin{figure}[h]
 \includegraphics[width=2.5cm]{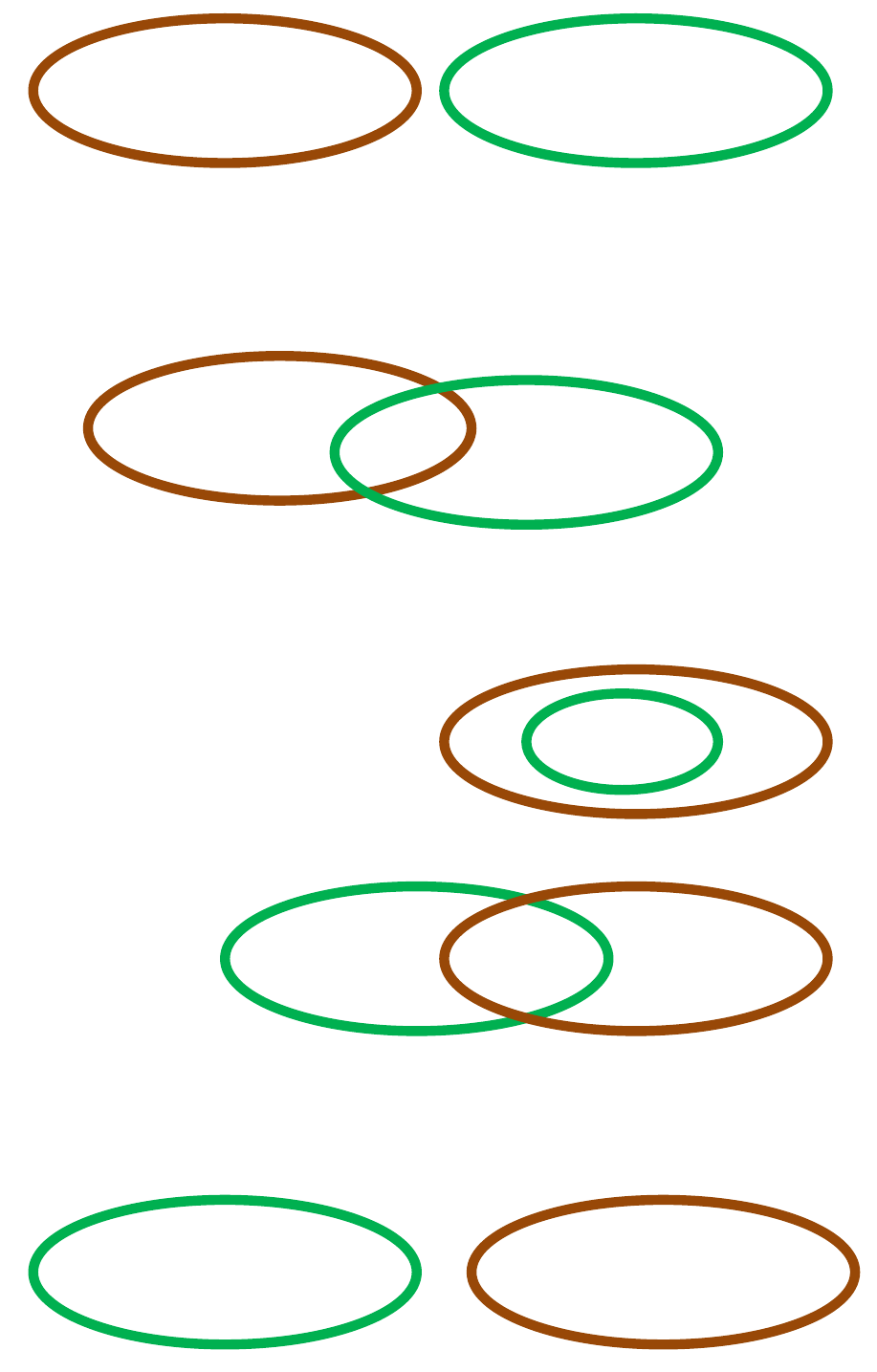} \hspace{2pc}
\begin{minipage}[b]{18pc} \caption{\label{fig10} Reading upwards, the left (green) loop passes under and through the right (brown) loop with the final positions of the two loop interchanged.}\vspace{0.4in}
\end{minipage}
\end{figure}
\end{center}

Can we generalize our model for 2-dimensional systems to 3-dimensions in a meaningful way? 
By the above-mentioned spin statistics theorem, point-like particles in 3 dimensions do not admit interesting braiding statistics.  However, the motions of \textit{loop-like} particles (e.g. vortices) in 3-dimensional space is mathematically interesting, and physical realizations are being studied as well \cite{LevinWang}.  Consider a collection of $n$ identical oriented loops (circles) inside a ball.  There are two obvious local symmetries
\textbf{ Loop interchange}\quad {\LARGE ${\color{green}\mathbf{\bigcirc}}\leftrightarrow{\color{brown}\mathbf{\bigcirc}}$} \quad and \textbf{Leapfrogging}, see Figure \ref{fig10}.
As we assume the loops are oriented we do not permit a flip of a single loop, since this reverses the orientation.
If we denote by $s_i$ the interchange of loops $i$ and $i+1$ and by $\sigma_i$ the leapfrogging operation on loops $i$ and $i+1$ the corresponding trajectories are in $3+1$-dimensional space-time.  We can visualize them as in Figure \ref{fig11}.
\begin{figure}[h]
\begin{tikzpicture}[scale=0.7,every node/.style={scale=0.7}]
      \node[] at ({-4*\rsep},{\maxy/2}) {\Large{$\sigma_{i} = $}};
      \begin{scope}[xshift = 0]
        \draw (0,0) arc (180:360:{\hrad} and {\vrad});
        \draw[dashed,color=black!80!white] (0,0) arc (180:0:{\hrad} and {\vrad});
        \draw (0,{\miny}) -- (0,{\maxy});
        \draw (1,{\miny}) -- (1,{\maxy});
        \draw (0,{\maxy}) arc(180:360:{\hrad} and {\vrad});
        \draw (0,{\maxy}) arc(180:0:{\hrad} and {\vrad});
        \node[] at ({\hrad},{\miny-2*\rsep}) {$1$};
      \end{scope}
      \node at (2,{\maxy/2}) {\Huge{$\cdots$}};
      \begin{scope}[xshift = 3cm]
        \draw plot [smooth,tension={\tension}] coordinates{
        ({\hrad-\hrad},0)
        ({\hrad-0.8*\hrad},{\maxy/5})
        ({2*\hrad+\rsep-0.8*\hrad},{2*\maxy/5})
        ({2*\hrad+\rsep-0.7*\hrad},{3*\maxy/5})
        ({3*\hrad+2*\rsep-1.15*\hrad},{4*\maxy/5})
        ({3*\hrad+2*\rsep-\hrad},{5*\maxy/5})};
        \draw plot [smooth,tension={\tension}] coordinates{
        ({\hrad+\hrad},0)
        ({\hrad+1.15*\hrad},{\maxy/5})
        ({2*\hrad+\rsep+0.7*\hrad},{2*\maxy/5})
        ({2*\hrad+\rsep+0.8*\hrad},{3*\maxy/5})
        ({3*\hrad+2*\rsep+0.8*\hrad},{4*\maxy/5})
        ({3*\hrad+2*\rsep+\hrad},{5*\maxy/5})};
        \draw[line width={60*0.8*\hrad},color=white] plot [smooth,tension={\tension}] coordinates{
        ({3*\hrad+2*\rsep},0)
        ({3*\hrad+2*\rsep},{\maxy/5})
        ({2*\hrad+\rsep},{2*\maxy/5})
        ({2*\hrad+\rsep},{3*\maxy/5})
        ({\hrad},{4*\maxy/5})
        ({\hrad},{5*\maxy/5})};
        \draw plot [smooth,tension={\tension}] coordinates{
        ({3*\hrad+2*\rsep-\hrad},0)
        ({3*\hrad+2*\rsep-1.1*\hrad},{\maxy/5})
        ({2*\hrad+\rsep-\hrad},{2*\maxy/5})
        ({2*\hrad+\rsep-1.1*\hrad},{3*\maxy/5})
        ({\hrad-0.9*\hrad},{4*\maxy/5})
        ({\hrad-\hrad},{5*\maxy/5})};
        \draw plot [smooth,tension={\tension}] coordinates{
        ({3*\hrad+2*\rsep+\hrad},0)
        ({3*\hrad+2*\rsep+0.9*\hrad},{\maxy/5})
        ({2*\hrad+\rsep+1.1*\hrad},{2*\maxy/5})
        ({2*\hrad+\rsep+\hrad},{3*\maxy/5})
        ({\hrad+1.1*\hrad},{4*\maxy/5})
        ({\hrad+\hrad},{5*\maxy/5})};
        \draw[dashed,color=black!80!white] plot [smooth,tension={\tension}] coordinates{
        ({\hrad-\hrad},0)
        ({\hrad-0.8*\hrad},{\maxy/5})
        ({2*\hrad+\rsep-0.8*\hrad},{2*\maxy/5})
        ({2*\hrad+\rsep-0.7*\hrad},{3*\maxy/5})
        ({3*\hrad+2*\rsep-1.15*\hrad},{4*\maxy/5})
        ({3*\hrad+2*\rsep-\hrad},{5*\maxy/5})};
        \draw[dashed, color=black!80!white] plot [smooth,tension={\tension}] coordinates{
        ({\hrad+\hrad},0)
        ({\hrad+1.15*\hrad},{\maxy/5})
        ({2*\hrad+\rsep+0.7*\hrad},{2*\maxy/5})
        ({2*\hrad+\rsep+0.8*\hrad},{3*\maxy/5})
        ({3*\hrad+2*\rsep+0.8*\hrad},{4*\maxy/5})
        ({3*\hrad+2*\rsep+\hrad},{5*\maxy/5})};
        \draw ({0.95*\hrad+\rsep},{2.5*\maxy/5}) arc (180:360:{1.05*\hrad} and {1.1*\vrad});
        \draw[dashed,color=black!20!white] ({0.95*\hrad+\rsep},{2.5*\maxy/5}) arc (180:0:{1.05*\hrad} and {1.1*\vrad});
        \draw[color=white] ({0.95*\hrad+\rsep},{2.5*\maxy/5}) arc (180:130:{1.05*\hrad} and {1.1*\vrad});
        \draw[dashed,color=black!80!white] ({0.95*\hrad+\rsep},{2.5*\maxy/5}) arc (180:130:{1.05*\hrad} and {1.1*\vrad});
        \draw[color=white] ({0.95*\hrad+\rsep+2.1*\hrad},{2.5*\maxy/5}) arc (0:50:{1.05*\hrad} and {1.1*\vrad});
        \draw[dashed,color=black!80!white] ({0.95*\hrad+\rsep+2.1*\hrad},{2.5*\maxy/5}) arc (0:50:{1.05*\hrad} and {1.1*\vrad});
        \draw[dashed,color=black!80!white] ({2*\hrad+\rsep-0.7*\hrad},{2.5*\maxy/5}) arc (180:360:{0.7*\hrad} and {0.5*\vrad});
        \draw[dashed,color=black!50!white] ({2*\hrad+\rsep-0.7*\hrad},{2.5*\maxy/5}) arc (180:0:{0.7*\hrad} and {0.5*\vrad});
        \draw (0,0) arc (180:360:{\hrad} and {\vrad});
        \draw[dashed,color=black!80!white] (0,0) arc (180:0:{\hrad} and {\vrad});
        \draw (0,{\maxy}) arc(180:360:{\hrad} and {\vrad});
        \draw (0,{\maxy}) arc(180:0:{\hrad} and {\vrad});
        \draw ({2*\hrad+2*\rsep},0) arc (180:360:{\hrad} and {\vrad});
        \draw[dashed,color=black!80!white] ({2*\hrad+2*\rsep},0) arc (180:0:{\hrad} and {\vrad});
        \draw ({2*\hrad+2*\rsep},{\maxy}) arc(180:360:{\hrad} and {\vrad});
        \draw ({2*\hrad+2*\rsep},{\maxy}) arc(180:0:{\hrad} and {\vrad});
        \node[] at ({\hrad},{\miny-2*\rsep}) {$i$};
        \node[] at ({2*\hrad+2*\rsep+\hrad},{\miny-2*\rsep}) {$i+1$};
      \end{scope}
      \node at (6.5-\rsep,{\maxy/2}) {\Huge{$\cdots$}};
      \begin{scope}[xshift = 7cm]
        \draw (0,0) arc (180:360:{\hrad} and {\vrad});
        \draw[dashed,color=black!80!white] (0,0) arc (180:0:{\hrad} and {\vrad});
        \draw (0,{\miny}) -- (0,{\maxy});
        \draw (1,{\miny}) -- (1,{\maxy});
        \draw (0,{\maxy}) arc(180:360:{\hrad} and {\vrad});
        \draw (0,{\maxy}) arc(180:0:{\hrad} and {\vrad});
        \node[] at ({\hrad},{\miny-2*\rsep}) {$n$};
      \end{scope}
    \end{tikzpicture}
    \;
    \begin{tikzpicture}[scale = 0.7,every node/.style={scale=0.7}]
      \node[] at ({-4*\rsep},{\maxy/2}) {\Large{$s_{i} = $}};
      \begin{scope}[xshift = 0]
        \draw (0,0) arc (180:360:{\hrad} and {\vrad});
        \draw[dashed,color=black!80!white] (0,0) arc (180:0:{\hrad} and {\vrad});
        \draw (0,{\miny}) -- (0,{\maxy});
        \draw (1,{\miny}) -- (1,{\maxy});
        \draw (0,{\maxy}) arc(180:360:{\hrad} and {\vrad});
        \draw (0,{\maxy}) arc(180:0:{\hrad} and {\vrad});
        \node[] at ({\hrad},{\miny-2*\rsep}) {$1$};
      \end{scope}
      \node at (2,{\maxy/2}) {\Huge{$\cdots$}};

      \begin{scope}[xshift = 5.5cm,xscale=-1]
        \draw plot [smooth,tension={\tension}] coordinates{
        ({3*\hrad+2*\rsep-\hrad},0)
        ({3*\hrad+2*\rsep-1.05*\hrad},{\maxy/4})
        ({2*\hrad+\rsep-\hrad},{\maxy/2})
        ({\hrad-0.8*\hrad},{3*\maxy/4})
        ({\hrad-\hrad},{\maxy})};
        \draw plot [smooth,tension={\tension}] coordinates{
        ({3*\hrad+2*\rsep+\hrad},0)
        ({3*\hrad+2*\rsep+0.8*\hrad},{\maxy/4})
        ({2*\hrad+\rsep+\hrad},{\maxy/2})
        ({\hrad+1.05*\hrad},{3*\maxy/4})
        ({\hrad+\hrad},{\maxy})};
        \draw[color=white,line width={60*0.55*\hrad}] plot [smooth,tension={\tension}] coordinates{({\hrad},0) ({\hrad},{\maxy/4}) ({2*\hrad+\rsep},{\maxy/2}) ({3*\hrad+2*\rsep},{3*\maxy/4}) ({3*\hrad+2*\rsep},{\maxy})};
        \draw plot [smooth,tension={\tension}] coordinates{
        ({\hrad-\hrad},0)
        ({\hrad-0.8*\hrad},{\maxy/4})
        ({2*\hrad+\rsep-\hrad},{\maxy/2})
        ({3*\hrad+2*\rsep-1.05*\hrad},{3*\maxy/4})
        ({3*\hrad+2*\rsep-\hrad},{\maxy})};
        \draw plot [smooth,tension={\tension}] coordinates{
        ({\hrad+\hrad},0)
        ({\hrad+1.05*\hrad},{\maxy/4})
        ({2*\hrad+\rsep+\hrad},{\maxy/2})
        ({3*\hrad+2*\rsep+0.8*\hrad},{3*\maxy/4})
        ({3*\hrad+2*\rsep+\hrad},{\maxy})};
        \draw[dashed,color=black!50!white] plot [smooth,tension={\tension}] coordinates{
        ({3*\hrad+2*\rsep-\hrad},0)
        ({3*\hrad+2*\rsep-1.05*\hrad},{\maxy/4})
        ({2*\hrad+\rsep-\hrad},{\maxy/2})
        ({\hrad-0.8*\hrad},{3*\maxy/4})
        ({\hrad-\hrad},{\maxy})};
        \draw[dashed,color=black!50!white] plot [smooth,tension={\tension}] coordinates{
        ({3*\hrad+2*\rsep+\hrad},0)
        ({3*\hrad+2*\rsep+0.8*\hrad},{\maxy/4})
        ({2*\hrad+\rsep+\hrad},{\maxy/2})
        ({\hrad+1.05*\hrad},{3*\maxy/4})
        ({\hrad+\hrad},{\maxy})};
        \draw (0,0) arc (180:360:{\hrad} and {\vrad});
        \draw[dashed,color=black!80!white] (0,0) arc (180:0:{\hrad} and {\vrad});
        \draw (0,{\maxy}) arc(180:360:{\hrad} and {\vrad});
        \draw (0,{\maxy}) arc(180:0:{\hrad} and {\vrad});
        \draw ({2*\hrad+2*\rsep},0) arc (180:360:{\hrad} and {\vrad});
        \draw[dashed,color=black!80!white] ({2*\hrad+2*\rsep},0) arc (180:0:{\hrad} and {\vrad});
        \draw ({2*\hrad+2*\rsep},{\maxy}) arc(180:360:{\hrad} and {\vrad});
        \draw ({2*\hrad+2*\rsep},{\maxy}) arc(180:0:{\hrad} and {\vrad});
      \end{scope}
        \node[] at ({3+\hrad},{\miny-2*\rsep}) {$i$};
        \node[] at ({3+2*\hrad+2*\rsep+\hrad},{\miny-2*\rsep}) {$i+1$};
      \node at (6.5-\rsep,{\maxy/2}) {\Huge{$\cdots$}};
      \begin{scope}[xshift = 7cm]
        \draw (0,0) arc (180:360:{\hrad} and {\vrad});
        \draw[dashed,color=black!80!white] (0,0) arc (180:0:{\hrad} and {\vrad});
        \draw (0,{\miny}) -- (0,{\maxy});
        \draw (1,{\miny}) -- (1,{\maxy});
        \draw (0,{\maxy}) arc(180:360:{\hrad} and {\vrad});
        \draw (0,{\maxy}) arc(180:0:{\hrad} and {\vrad});
        \node[] at ({\hrad},{\miny-2*\rsep}) {$n$};
      \end{scope}
    \end{tikzpicture}
 \caption{\label{fig11}}

\end{figure}
The group generated by $\sigma_i$ and $s_i$ for $1\leq i\leq n-1$ is called the \textbf{Loop Braid Group}, $\mathcal{LB}_n$, defined abstractly as the group satisfying:

Braid relations:
\begin{enumerate}
 \item[(R1)]$\sigma_i\sigma_{i+1}\sigma_{i}=
\sigma_{i+1}\sigma_{i}\sigma_{i+1}$
\item[(R2)] $\sigma_i\sigma_j=\sigma_j\sigma_i$
if $|i-j|>1$
\end{enumerate}
Symmetric Group relations:
\begin{enumerate}
 \item[(S1)]$s_is_{i+1}s_{i}=
s_{i+1}s_{i}s_{i+1}$
\item[(S2)] $s_is_j=s_js_i$
if $|i-j|>1$
\item[(S3)] $s_i^2=1$
\end{enumerate}
Mixed relations:
\begin{enumerate}
 \item[(M1)]$\sigma_i\sigma_{i+1}s_{i}=
s_{i+1}\sigma_{i}\sigma_{i+1}$
\item[(M2)] $s_is_{i+1}\sigma_{i}=
\sigma_{i+1}s_{i}s_{i+1}$
\item[(M3)] $\sigma_is_j=s_j\sigma_i$
if $|i-j|>1$
\end{enumerate}

This is a relatively new area of development, for which many questions and research directions remain unexplored.  A first mathematical step is to study the unitary representations of the loop braid group, which is already underway \cite{loopleeds,mrcloop}.  It might also be reasonable to consider other configurations, such loops bound concentrically to an auxillary loop or knotted loops.

\subsection{Conclusions}
We have briefly illustrated how modeling the physical properties and computational applications of anyons on surfaces leads to a rich mathematical theory. This theory, in turn, can be used to probe fundamental questions and guide experiments in 2-dimensional topological phases of matter.  Moreover, topological considerations suggest that 3-dimensional materials might also be studied in an analogous way, using loop-like excitations.

\ack{
This article is based upon two lectures given at QuantumFest 2015 held at Monterrey Tec, Estado de Mexico campus.  I would like to thank that institution and the organizers for a stimulating conference and wonderful hospitality.  The author was partially supported by NSF grants.

 This article is dedicated to the memory of our friend Sujeev
Wickramasekara.}

\section{References}

\end{document}